# Prediction of Blood Lactate Values in Critically Ill Patients: A Retrospective Multi-center Cohort Study

**Authors**: Behrooz Mamandipoor[1], Wesley Yeung[2,3], Louis Agha-Mir-Salim[2,4], David J. Stone[5], Venet Osmani[1*], Leo Anthony Celi[2,6,7]

**Affiliations**:

[1]Fondazione Bruno Kessler Research Institute, Trento, Italy

[2]Laboratory for Computational Physiology, Harvard-MIT Health Sciences and Technology, Massachusetts Institute of Technology, Cambridge, MA, 02139, USA

[3]University Medicine Cluster, National University Hospital, Singapore

[4]Faculty of Medicine, University of Southampton, 12 University Rd, Southampton SO17 1BJ, United Kingdom

[5]Departments of Anesthesiology and Neurosurgery, and the Center for Advanced Medical Analytics, University of Virginia School of Medicine, Charlottesville, VA, 22908, USA

[6]Division of Pulmonary, Critical Care and Sleep Medicine, Beth Israel Deaconess Medical Center, Boston, MA 02215, USA

[7]Department of Biostatistics, Harvard T.H. Chan School of Public Health, Boston, MA 02115, USA

**\*Corresponding Author**
vosmani@fbk.eu

**Abstract**

**Purpose**. Elevations in initially obtained serum lactate levels are strong predictors of mortality in critically ill patients. Identifying patients whose serum lactate levels are more likely to increase can alert physicians to intensify care and guide them in the frequency of tending the blood test. We investigate whether machine learning models can predict subsequent serum lactate changes.

**Methods.** We investigated serum lactate change prediction using the MIMIC-III and eICU-CRD datasets in internal as well as external validation of the eICU cohort on the MIMIC-III cohort. Three subgroups were defined based on the initial lactate levels: i) normal group (<2 mmol/L), ii) mild group (2-4 mmol/L), and iii) severe group (>4 mmol/L). Outcomes were defined based on increase or decrease of serum lactate levels between the groups. We also performed sensitivity analysis by defining the outcome as lactate change of >10% and furthermore investigated the influence of the time interval between subsequent lactate measurements on predictive performance.

**Results.** The LSTM models were able to predict deterioration of serum lactate values of MIMIC-III patients with an AUC of 0.77 (95% CI 0.762-0.771) for the normal group, 0.77 (95% CI 0.768-0.772) for the mild group, and 0.85 (95% CI 0.840-0.851) for the severe group, with a slightly lower performance in the external validation.

**Conclusion.** The LSTM demonstrated good discrimination of patients who had deterioration in serum lactate levels. Clinical studies are needed to evaluate whether utilization of a clinical decision support tool based on these results could positively impact decision-making and patient outcomes.

**Keywords**: resuscitation, lactate, critical illness, deep learning, time series

# Introduction

Hyperlactatemia is a medical condition caused by accumulation of lactate and hydrogen ions in the bloodstream and tissues, usually as a result of tissue hypoxia and systemic hypoperfusion. It is commonly observed and treated in critical care conditions such as severe heart failure, sepsis, or other forms of shock. Both the magnitude and rate of change of serum lactate elevation are strong predictors of mortality [1].

When a patient is admitted to an ICU for certain conditions such as shock or trauma, a serum lactate level may be obtained in addition to a number of other laboratory tests. Certain tests including metabolic and hematologic (i.e. complete blood count) panels are routinely obtained on a daily basis during the acute phase of critical illness. However, it is rarely appropriate or indicated to test serum lactate in this kind of routine, periodic basis. Rather, serum lactate values are obtained in a targeted fashion based on the clinical context, or more explicitly, on the perceived stability of the patient. Reliance on the provider to order the test introduces variation that can impact the outcome of the patient. In this work, we address the issue of whether the clinical determination of timing for subsequent serum lactate samples can be improved by application of artificial intelligence techniques to available data.

Blood (serum, or plasma when the lactate is measured in an anticoagulated sample with an arterial blood gas) lactate reduction during the initial hours of intensive care unit (ICU) admission has been shown to be associated with improved survival [2]–[5], while persistently high and increasing levels are associated with poor outcomes [6]. Resuscitation guided by serum lactate levels has also been shown to be associated with reduced hospital mortality [7]. At present, continuous lactate monitoring is not yet available [8]. While frequent, periodic serum lactate measurements might seem the next best choice, these approaches involve downsides, including risk of anemia from repeated blood draws [9], [10], need for frequent venipunctures, or use of a central venous catheter to draw blood that comes with an infection risk [11], [12], and cost. Many unnecessary samples are also likely to be drawn when periodic studies are ordered. Most importantly, in the absence of the availability of continuous serum lactate measurements, the optimal approach to periodic or repeated determination of serum lactate level simply remains uncertain, and in current practice is likely to rest on the variable and individualized experiences of practitioners as well as the inevitable exigencies and vicissitudes of clinical workflow in a demanding environment. The fundamental clinical question is whether the serum lactate levels are likely to be increasing, or in the case of already elevated levels, showing no improvement (which is considered negatively from a clinical standpoint). A data driven trigger for determining the need and timing for repeat serum lactate testing would be a significant advance in standardizing and potentially improving care processes as well as clinical laboratory utilization in this setting.

Given the strong prognostic utility of serum lactate [1], continuously predicting the trajectory of serum lactate values would be clinically useful as a tool that would optimize the number of serum lactate tests by reducing unnecessary testing while providing a reminder for necessary, and presumably useful, subsequent testing. A prediction of increasing serum lactate levels could alert clinicians to potential deterioration and prompt confirmatory testing with a blood draw. On the other hand, a prediction of stable (in the case of previously normal levels) or improving serum lactate levels would prevent unnecessary blood draws. Machine learning algorithms may be useful in both prompting repeat blood draws likely to yield actionable information, and in reducing the number of unnecessary repeat testing [13].

We hypothesize that: 1) clinical variables during the first 48 hours of ICU admission can predict the trajectory of serum lactate values during that time, and that 2) patients classified into normal, mild and severe groups, based on their initial serum lactate measurements, manifest different factors affecting this trajectory. In this work, we describe an approach to detecting worsening hyperlactatemia in ICU patients on the basis of input of expert clinical knowledge, state-of-the-art analytical techniques, and large, high-resolution, multi-center datasets to construct three models to identify patients at risk of worsening hyperlactatemia within the first 48 hours of ICU admission.

# Methods

## Data Sources

The Medical Information Mart for Intensive Care (MIMIC-III, v1.4) is a longitudinal, single-center database maintained by the Laboratory for Computational Physiology at the Massachusetts Institute of Technology (MIT) which contains data associated with 53,423 distinct ICU admissions for adult patients (aged 16 years and older) admitted to critical care units between 2001 and 2012 [14] at the Beth Israel Deaconess Medical Center. It is a teaching hospital of Harvard Medical School with 673 licensed beds, including 493 medical/surgical beds, 77 critical care beds, and 62 OB/GYN beds.
The eICU Collaborative Research Database (eICU-CRD) contains data associated with 200,859 admissions collected from 335 ICUs across 208 hospitals in the US admitted between 2014 and 2015 [15].

## Study Design

We retrospectively evaluated a subgroup of adult patients (age ≥ 18 years) from the MIMIC-III and eICU-CRD datasets that had at least 2 serum lactate measurements recorded within the first 48 hours of ICU admission as well as an ICU length of stay greater than or equal to 24 hours. The selected patients were further divided into three subgroups based on their initial serum lactate levels: i) normal group (< 2 mmol/L), ii) mild group (2 mmol/L to 4 mmol/L) and iii) severe group (> 4 mmol/L). The present study is reported in accordance with the Strengthening the Reporting of Observational studies in Epidemiology (STROBE) statement.

## Definition of Outcomes

The outcome was the trajectory of the second serum lactate measurement which was categorized into a positive or negative outcome based on the initial subgroup to which the patient belonged. For the normal group, a negative outcome was defined as a serum lactate increase to mild or severe levels, while a positive outcome was defined as a value that remained within the normal level. A similar approach was taken for the other two subgroups, as shown in Table 1, where the increase in lactate levels between groups corresponds to a negative outcome.

*Table 1 Definition of outcomes for each patient subgroup*

|  | Outcome | |
| --- | --- | --- |
| **Initial Lactate Value** | Negative | Positive |
| Normal (< 2 mmol/L) | Serum lactate **increases** to mild or severe group levels | Serum lactate **remains** within the normal group |
| Mild (2 mmol/L to 4 mmol/L) | Serum lactate **increases** to severe group or **remains** within the mild group levels | Serum lactate **decreases** to normal group levels |
| Severe (> 4 mmol/L) | Serum lactate **increases** or **remains** within the severe group levels | Serum lactate **decreases** to mild or normal group levels |

## Sensitivity Analyses

We conducted a sensitivity analysis to investigate whether a 10% change in serum lactate levels (rather than between groups) influences predictive performance of the model. The 10% change was chosen because serum lactate non-clearance, defined as a serum lactate decrease of less than 10%, is associated with an increased risk of mortality [2]–[6]. Details and results of this analysis are presented in Appendix 1.

We also conducted an additional sensitivity analysis to investigate whether the difference in time between the two serum lactate measurements has any effect on the prediction performance of lactate deterioration. For this analysis, we restricted the cohort to only those patients that had the subsequent serum lactate measured within 8 hours of the preceding lactate measurement. The 8-hour interval was chosen based on Surviving Sepsis Campaign guidelines [16] that recommend serum lactate be measured every 6 hours. We allowed for a 2-hour delay to account for situations where the serum lactate might be measured but not immediately recorded in the patient's health record. Details and results of these analyses are presented in Appendix 2.

## Variable selection

We selected 54 variables identified by ICU clinicians and the related literature as relevant to serum lactate deterioration and available in both MIMIC-III and eICU-CRD within 48 hours of admission. These include other selected laboratory values, vital signs, patient demographics, and nursing care data obtained during the admission assessment as shown in Appendix 3. These also included values obtained through an arterial blood gas (ABG) measurement which had to be sampled at least two hours (discretization interval) before the timestamp of the predicted serum lactate level. Laboratory variables were discretized into two-hour intervals as experiments revealed better model performance compared to models developed on hourly time windows. Outliers were addressed by defining a clinically valid interval. The variables were normalized using zero mean and scaling to unit variance. Linear correlation (Pearson) between the top 10 highest correlated variables with serum lactate is shown in Appendix 3.

## Missing Values Imputation Strategy

We evaluated several imputation strategies using both data-driven approaches and in combination with clinical heuristics. In our previous work, we evaluated twelve different imputation strategies, including strategies based on mean, multiple imputation (chained equations), random forest, and autoencoders for prediction of serum lactate levels [17], [18]. This previous work has shown that an imputation strategy based on mean and indicator variables, where a Boolean variable is added to indicate whether a value is missing or not, provides the best performance. We therefore set out to evaluate this strategy first, and subsequently compare it with the most common strategy based on the mean value. The results showed that the indicator imputation strategy provided better performance (in terms of AUC) than using the mean value alone. However, using an indicator variable degrades model interpretability and variable ranking, due to the increase in the number of Boolean variables (an indicator is added for each variable with missing values).

As such, we opted for a fill-forward imputation strategy applied to each ICU stay by forward propagation of all the valid measurements. This approach provided an optimal trade-off between model performance and interpretability. Furthermore, we investigated whether a strategy based on clinical heuristics would further improve performance. Using this strategy, we defined an imputation method for each individual variable, as shown in Appendix 3. This strategy provided 2% (±0.85) AUC better performance on average than using the mean value. As a result, after splitting datasets into train and test set, we used clinical heuristics combined with the fill-forward method for the imputation of missing values. It should be noted that no lactate values were imputed.

# Experimental evaluation methodology

### Model development and experimentation
We evaluated the performance of three machine learning algorithms - logistic regression (LR), random forest (RF), and long short-term memory (LSTM). LR is an algorithm capable of predicting class probabilities using predictor variables, by adjusting the coefficients of the logit function, RF is an ensemble learning method constructing multiple decision trees and then producing class probabilities as outputs [19], while LSTM is a type of Deep Artificial Neural Network designed to learn temporal dependencies between variables and process longitudinal time-series data [20].

We spilt the data randomly into a derivation cohort (80%) and validation cohort (20%), where hyperparameters of all the models were optimized using random search on the validation set, detailed in Appendix 4. The final models were internally validated using stratified five-fold cross validation with 5 repetitions for both MIMIC-III and eICU-CRD datasets. For the external validation, we derived models on the eICU-CRD patient cohort and validated them on the MIMIC-III cohort.

### Performance evaluation
We assessed each model by computing the area under the receiver operator characteristic curve (AUC-ROCs) and the area under the precision-recall curve (AUPRCs), also called Average Precision (AP). We also provide additional performance metrics, including calibration, Positive Predictive Value (PPV), Negative Predictive Value (NPV), F-1 score (showing the balance between PPV and sensitivity) and Matthews correlation coefficient MCC (used to measure the quality of classification between our algorithms) for each model. These performance metrics are detailed in Appendix 5, whereas calibration performance is detailed in Appendix 6.

### Model interpretability
We conducted a model interpretability analysis to understand how the model ranked the importance of variables when predicting serum lactate trajectory. We used the SHAP (Shapley Additive exPlanations) method whose objective is to explain a prediction output of a machine learning model by computing the contribution of each variable to the prediction [21]. The SHAP method computes Shapley values, where those variables with the largest absolute values are the most important. Based on Shapley values, we ranked each variable based on importance, including a ranking of the top ten variables.

# Results

## MIMIC-III cohort

From the 61,532 overall admissions in MIMIC-III, 12,502 admissions matched our selection criteria (11,083 patients). The cohort selection diagram is shown in Appendix 3.

The MIMIC-III cohort had 29,337 serum lactate values recorded within the first 48 hours of ICU admission with close to half in the normal group (46.9%) with the remaining values in the mild (37.7%) and severe groups (15.4%). The average patient age was 64.4 (±16.6) with 42% female patients. The most common admission diagnosis was sepsis, followed by pneumonia, with a median length of stay of 4.2 (± 8.6) days and mortality rate of 20.7% as shown in Table 2. Detailed patient characteristics and subgroup differences for both MIMIC-III and eICU-CRD are shown in Appendix 3.

*Table 2 MIMIC-III cohort characteristics based on the initial lactate measurement*

|  | **Overall** | **Normal** | **Mild** | **Severe** |
|---|---|---|---|---|
| **Age (mean)** | 64.4 (± 16.6) | 64.6 (± 16.5) | 64.4 (± 16.7) | 62.8 (± 17.0) |
| **Gender (m)** | 7211 (58%) | 4502 (57%) | 3777 (58%) | 1380 (58%) |
| **Length of Stay (median)** | 4.2 (± 8.6) | 4.7 (± 8.6) | 4.5 (± 8.9) | 5.1 (± 9.5) |
| **Mortality** | 20.7 % | 16.6 % | 21.5 % | 37.9 % |
| **Admission diagnosis** | Sepsis (5.0%) Pneumonia (3.8%) | Sepsis (4.6%) Pneumonia (4.3%) | Sepsis (5.3%) Pneumonia (3.0%) | Sepsis (5.9%) Abdominal Pain (2.2%) |

For the MIMIC-III cohort, positive outcomes (see Table 1) with respect to serum lactate trajectory were observed in 87.1% (n=11,977) of subsequent lactate measurements in the normal group, 40.5% (n=4,485) in the mild group, and 39.5% (n=1785) in the severe group. Figure 1 summarizes the performance of each model for each subgroup. We have also calculated precision-recall for different thresholds to devise curves that compare the performance of the models in the presence of imbalanced datasets, as is the case with the normal group where positive outcomes (87.1%) significantly exceed negative outcomes (12.9%).

RF and LR performed similarly for the normal group with AUCs for both of 0.74 (95% CI 0.738-0.748 vs 0.732-740), while RF outperformed LR for the other two subgroups. The LSTM model performed best across all the three subgroups, achieving an AUC of 0.77 (95% CI 0.762-0.771) for the normal group, 0.77 (95% CI 0.768-0.772) for the mild group, and 0.85 (95% CI 0.840-0.851) for the severe group. The models were well calibrated for the mild and severe groups as shown in Appendix 6, while the normal group demonstrated less accurate calibration.

We also investigated the performance of the models when removing the patients with elective admission type. However, the results showed no statistically significant difference between the overall cohort and the cohort without elective admissions in terms of AUC performance.

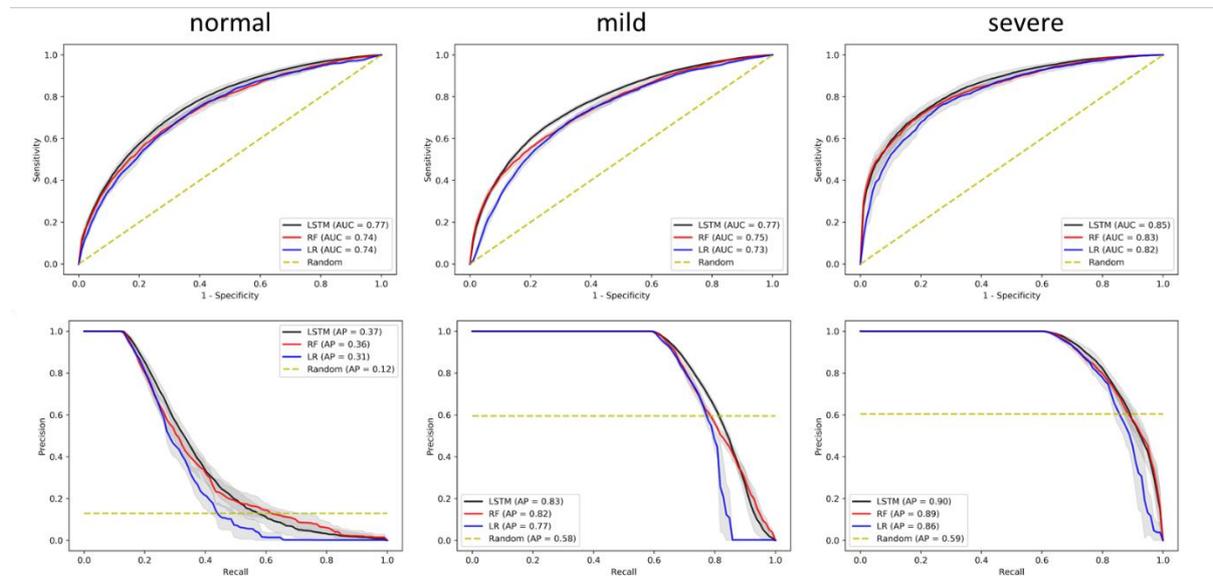

*Figure 1 Performance of each model in the MIMIC-III cohort across the three patient subgroups. Top row represents AUC-ROC, while the bottom AU-PRC. Confidence intervals are shown in grey.*

## eICU-CRD Cohort

From the 200,859 admissions in the eICU-CRD, 17,452 admissions (16,283 patients) matched our selection criteria, as detailed in the cohort selection diagram in Appendix 3. The eICU-CRD study cohort had 39,389 serum lactate values recorded within the first 48 hours of ICU admission with 39.7% in the normal group, 35.4% in the mild group, and 24.9% in the severe group. The average patient age was 62.2 (± 16.1) with 45% female patients. The most common admission diagnosis was sepsis, followed by cardiac arrest, with a median length of stay of 3.5 (± 6.5) days and mortality rate of 15.4% as shown in Table 3.

*Table 3 eICU-CRD cohort characteristics*

|  | **Overall** | **Normal** | **Mild** | **Severe** |
|---|---|---|---|---|
| **Age (mean)** | 62.2 (± 16.1) | 61.9 (± 16.0) | 62.3 (± 16.3) | 61.8 (± 16.0) |
| **Gender (m)** | 9520 (55%) | 5142 (54%) | 4772 (56%) | 2649 (55%) |
| **Length of Stay (median)** | 3.5 (± 6.5) | 3.7 (± 6.3) | 3.7 (± 6.6) | 3.8 (± 7.0) |
| **Mortality** | 15.4 % | 10.0 % | 15.3 % | 29.4 % |
| **Admission diagnosis** | Sepsis, pulmonary (12.7%) Cardiac arrest (8.1%) | Sepsis, pulmonary (12.8%) Cardiac arrest (7.5%) | Sepsis, pulmonary (13.3%) Cardiac arrest (8.7%) | Cardiac arrest (15.0%) Sepsis, pulmonary (11.3%) |

For the eICU-CRD cohort, positive outcomes with respect to serum lactate trajectory were observed in 87.2% (n=13,640) in the normal group, 40.4% (n=5,638) in the mild group, and 36.0% (n= 3,528) in the severe group. Figure 2 summarizes the performance of each model for each subgroup. RF performed slightly better than LR for the mild group with an AUC of 0.73 (95% CI 0.723-0.731) versus an AUC of 0.69 (95% CI 0.683-0.693), while both models had similar performances for the other two subgroups. The LSTM model performed best across all the three subgroups, achieving an AUC of 0.72 (95% CI 0.707-0.724) for the normal group, 0.74 (95% CI 0.735-0.745) for the mild group, and 0.84 (95% CI 0.837-0.848) for the severe group.

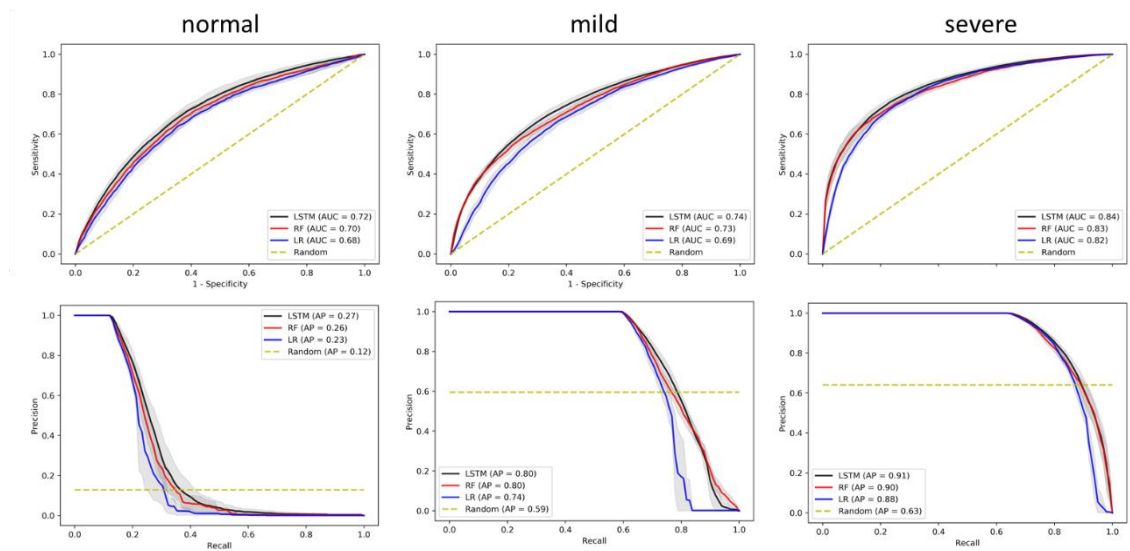

*Figure 2 Performance of each model in the eICU-CRD cohort across the three patient subgroups. Top row represents AUC-ROC, while the bottom AU-PRC. Confidence intervals are shown in grey.*

## External validation of the eICU-CRD model on the MIMIC-III cohort

In addition to evaluating serum lactate deterioration prediction within MIMIC-III and eICU-CRD individually, we conducted an external validation where a model derived from the eICU-CRD was validated on the MIMIC-III patient cohort. This was done to investigate the generalizability of our method on independent patient data and its potential utility as a clinical decision support tool. We followed the same cohort selection criteria and derived the model using the eICU-CRD patient cohort, while the MIMIC-III cohort was used as a test set. The results are detailed in Figure 3.

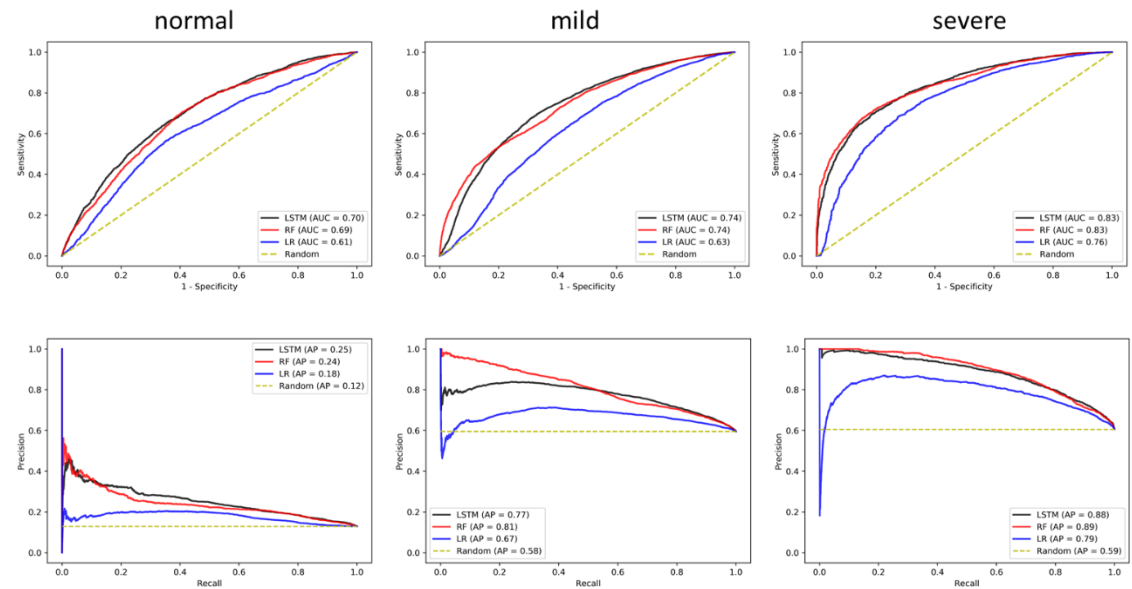

*Figure 3 Performance of each model derived in the eICU-CRD cohort and externally validated on the MIMIC-III cohort across the three patient subgroups. Top row represents AUC-ROC, while the bottom AU-PRC.*

It is typically much more challenging to achieve similar results with external validation in comparison to internal validation. However, the performance of our method remained at similar levels to the internal studies. External validation results demonstrated deepened differences in performance between baseline algorithms (LR) and machine learning approaches (LSTM and RF): the performance of LSTM was much closer to that of RF, outperforming it only in the normal group. While, in contrast to RF, the LSTM is equipped to capture temporal dependencies between serum lactate measurements, the majority of temporal sequences are quite short, especially in the severe group where serum lactate measurements are more frequent (see distribution of timing between subsequent lactate measurements in Appendix 3). LSTM and RF achieved AUCs of 0.74 and 0.83 for the mild and severe groups respectively, with a lower performance for the normal group with AUCs of 0.70 and 0.69.

## Sensitivity Analyses

We also conducted sensitivity analyses using an alternative definition of serum lactate prediction outcome where we defined a change of at least 10% of serum lactate levels as an increase or decrease. The results of this analysis are detailed in Appendix 1, where the AUC of the LSTM model decreased to 0.76 (95% CI 0.755-0.766) from 0.77 (95% CI 0.762-0.771) for the normal group; to 0.67 (95% CI 0.661-0.672) from 0.77 (95% CI 0.768-0.772) for the mild group; and to 0.75 (95% CI 0.746-0.758) from 0.85 (95% CI 0.840-0.851) for the severe group.

The second sensitivity analysis focused on investigating whether the time difference between subsequent serum lactate measurements has any effect on serum lactate deterioration prediction performance (detailed in Appendix 2). The results of this analysis showed a decrease in AUC performance of the LSTM model for the mild and severe group to an AUC of 0.75 (95% CI 0.744-0.751) from 0.77 (95% CI 0.768-0.772); and to 0.83 (95% CI 0.831-0.840) from 0.85 (95% CI 0.840-0.851), respectively. For the normal group, the performance increased slightly to an AUC of 0.78 (95% CI 0.772-0.783) from 0.77 (95% CI 0.762-0.771). These changes are not statistically significant.

## Variable importance

As we derived three different models for each of the subgroups, we also calculated variable rankings separately for each model. Therefore, the top three ranked variables for the model of the normal group were the prior serum lactate, serum glucose, and anion gap; for the mild group model- prior serum lactate, respiratory rate, and serum glucose; while, for the model of the severe group, the most important variables were prior serum lactate, prior arterial base excess, and Glasgow Coma Score value. A graphical representation of the top 10 variables for each model and their ranking is provided in Figure 4.

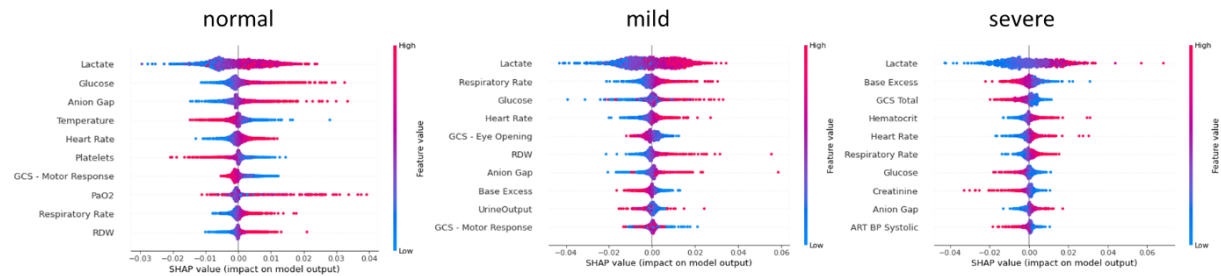

*Figure 4 Variable importance ranking for each LSTM-based model of the patient subgroups derived from the MIMIC-III cohort .*

## Discussion

The LSTM models were the most accurate in predicting deterioration of serum lactate values in all three serum lactate level subgroups in the MIMIC-III cohort, with an AUC of 0.77 (95% CI 0.762-0.771) for the normal group, 0.77 (95% CI 0.768-0.772) for the mild group, and 0.85 (95% CI 0.840-0.851) for the severe group.

We observed different patterns of importance of the variables among the patient subgroups. For example, in the subgroup with normal baseline serum lactate levels, the prior serum lactate measurement was an important predictor of deterioration of serum lactate values, followed by serum glucose, anion gap, temperature and heart rate. The mild and severe groups additionally showed respiratory rate, GCS, and base excess as important variables in predicting serum lactate levels. Arterial pH, base excess, serum bicarbonate, and serum anion gap values all reflect the acid-base balance [22], while decreased partial pressure of carbon dioxide and increased respiratory rate are the results of physiologic responses to metabolic acidosis [23]. Urine output is a function of volume status, and renal function, and is affected by vascular perfusion to that organ, which in turn is intrinsically linked with tissue acidosis and lactate metabolism [24][25]. Heart rate can be affected by many diverse factors and has been found to be independently associated with in-hospital mortality [26]. Elevated serum bilirubin could be a marker of hepatic metabolic dysfunction, commonly referred to as "shock liver", which has also been found to be an important predictor of survival [27]. Moreover, lactate is also metabolized by the liver so that hepatic dysfunction can independently contribute to worsening hyperlactatemia [28], [29]. Clearly, it makes sense that these values would be and are strong determinants of the ongoing state of lactate levels. In addition to forming the basis of our predictive models, knowing the relative weights of these values in contributing to observed lactate levels is also useful to know for clinicians making these decisions. While severe changes in blood pressure or pH are rather obvious indicators that another lactate value should be obtained, there is also a more subtle constellation of changes in laboratory and vital sign values that should drive clinicians to consider rechecking lactate levels.

Clinical decision support (CDS) modalities must be accurate, useful, and usable, and fit as seamlessly as possible into clinicians' workflows. Lindsell and co-workers [30] recently stated that "Designing a useful AI tool in health care should begin with asking what system change the AI tool is expected to precipitate." In this case, the change would consist of implementing a tool that would optimize the number of serum lactate tests by reducing unnecessary testing while providing a reminder for necessary, and presumably useful, subsequent testing.

The net change in test frequency would not be an adequate metric for evaluating the impact of this process change because any decrease in unnecessary testing could be offset by an increase in indicated testing. One metric that could be employed would be the relative (compared to baseline) percentage of repeat serum lactate values that demonstrated values that were clinically actionable (e.g., crossing the threshold from normal to mild, or from mild to high). But the critical metric would be whether the more focused identification of serum lactate anomalies contributes to improved outcomes in those patients who, at some time in their clinical course, have elevated serum lactate levels of some degree. The ultimate analysis of the value of such a CDS tool really requires a systems level approach that incorporates the classic ICU metrics of mortality and LOS, but also considers costs, fluid balance and renal function, impacts on workflows, and even the detection of adverse event outlier cases where the CDS leads the clinicians astray.

We would envision that the preliminary version of a CDS model would be updated every hour in order to make predictions employing events and values, newly captured over that interval. The algorithm would incorporate all pertinent variables in the interval and determine which contribute to the possible need for an additional serum lactate sample, e.g., new evidence of sepsis (e.g. increasing SOFA score), increasing anion gap, or increasing respiratory rate. The specific element of 'figuring out' the new determination only applies to the LSTM based model in which newly available data can actually update the state of the neural network. Unlike the static algorithms of LR and RF, the LSTM algorithm would change dynamically during use so that any CDS tool based on this approach would require special Food and Drug Administration (FDA) approval. The FDA is currently studying what will be best practices in terms of approving such innovative yet evolving instruments in clinical care. Further advances in the quality of the CDS would involve such dynamically evolving tools that learn continuously and provide automatic feedback to improve the model; learning how to best incorporate continuously tracked values such as HR and RR; identifying patient, disease, and unit level characteristics that could make the CDS a more precision tool; and the addition of new input variables as clinical medicine evolves.

Since all available data up to the time of prediction would be employed, even the fixed algorithms (LR, RF) would become potentially more accurate over time. For example, if there are two (or more) prior serum lactate values entered in the laboratory information system, then subsequent predictions should benefit from the creation of a potentially more robust trajectory than if only a single prior lactate is available (as is the case for this paper). Potentially, the AUCs generated in this instance would be at least as high, and likely higher than those we have reported, making the tool a progressively more accurate and useful one.

The relatively suboptimal model performance (AUC 0.72 – 0.83) compared to other machine learning in healthcare publications (including ours), highlights the challenge of predicting the trajectory of serum lactate during critical illness. Features pertaining to the immunologic response that are specific to a patient, likely are only partially captured in the clinical data currently collected in the process of care. However, we assert that having a model with decent discrimination to inform clinicians when to check serum lactate is still an improvement compared to the variation across clinicians with regard to when serum lactate is ordered. We suspect the model performance can be improved by training on a larger dataset. An acceptable precision should be set if the intent is to reduce unnecessary testing and by how much. An acceptable recall should be set if the intent is to detect deterioration early and increase patient monitoring or move the patient to a higher level of care.

While our results are specifically calculated for serum lactate, the method could, upon appropriate contextual recalculation of the algorithm, be applied to other often repeated laboratory tests such as serum glucose or hemoglobin, similarly alerting clinicians of the need to recheck a value on the basis of a predicted high probability of a clinically important and potentially actionable change in that value.

Our results suggest that it is feasible to predict future deterioration of serum lactate levels using routine clinical observations. However, there are several limitations. Validation studies on a target population are necessary before these models can safely be deployed in a clinical setting. The models also require regular recalibration to capture shifts in clinical practice and patient profiles over time. With these safeguards in place, our models could serve as a point-of-care tool that assists in the prediction of serum lactate values. This could provide an early warning for potential deterioration, prompting confirmatory testing of serum lactate levels (potentially automatically in the future) while at the same time reducing the number of unnecessary serum lactate blood tests for stable patients, and enable clinicians to have a more personalized approach to the care of critically ill patients. The approach also allows for clinicians to independently determine the need for serum lactate values, so that it represents a supplement to, rather than a replacement of, clinical judgment.

To the best of our knowledge, this is the first study that attempts to use physiological and routinely measured markers to predict serum lactate deterioration. Strengths of this study include robust machine learning methodology able to capture temporal relationships between time-varying variables, while at the same time providing interpretability of the results. However, as this was a retrospective study, there were missing data for some of the variables. While our external validation results are encouraging, the models trained on the MIMIC-III data were obtained from a single center in the United States, and its performance might not generalize to external populations; the eICU-CRD data were restricted to ICUs in the United States and might not generalize to global populations. Our intention with this study was to provide insight into appropriate methodology and present a sound approach in predicting a biochemical marker for clinical application as opposed to a highly accurate ungeneralizable model.

Clinical prediction models demonstrate their true utility only if they can positively impact clinical practice and patient outcomes without unacceptably negative impacts on workflow and/or costs. Apart from validation studies as described above, prospective evaluation in a clinical setting is required to measure the impact such models have on clinical decision making by nurses and physicians, and whether the subsequent changes in practice translate to desirable outcomes related to number of serum lactate tests ordered, rates of organ failure, length of ICU stay, and hospital mortality. We provide the parameters used to develop the models in the supplementary information to enable replication and extension of this work in other patient populations.

Conclusion
Compared to other clinical outcome prediction algorithms, our model performance seems suboptimal. Serum lactate is challenging to predict as lactate metabolism results from a complex interplay of factors pertaining to the patient, factors pertaining to the disease or injury, treatments, and patient response to treatment, and perhaps a signature genetically encoded host response. Omics data may be a proxy of the last element but is currently not captured in routine patient care. The difficulty with serum lactate prediction is in effect a missing data issue. Despite this, the LSTM model provided the highest performance in predicting lactate value deterioration in critically ill patients, followed by RF and LR. This suggests that the use of machine learning might be a useful adjunct in helping to predict serum lactate deterioration in a manner that can inform clinician decision-making. Further studies are needed to evaluate its utility in clinical practice.

## List of abbreviations

| | |
|---|---|
| ABG | Arterial Blood Gas |
| AI | Artificial Intelligence |
| AP | Average Precision |
| AUC | Area Under the Receiver Operating Characteristic Curve |
| AUPRC | Area Under the Precision Recall Curve |
| FDA | Food and Drug Administration |
| ICU | Intensive Care Unit |
| LR | Logistic Regression |
| LSTM | Long short-term memory |
| MCC | Mathews Correlation Coefficient |
| MIMIC | Medical Information Mart for Intensive Care |
| NPV | Negative Predictive Value |
| PPV | Positive Predictive Value |
| PRC | Precision Recall Curve |
| RF | Random Forest |
| ROC | Receiver Operating Characteristic |
| SHAP | SHapley Additive exPlanations |
| SOFA | Sequential Organ Failure Assessment |
| STROBE | STrengthening the Reporting of OBservational studies in Epidemiology |

# Declarations

**Ethical approval and consent to participate**

The data in MIMIC-III was previously de-identified, and the institutional review boards of the Massachusetts Institute of Technology (No. 0403000206) and Beth Israel Deaconess Medical Center (2001-P-001699/14) both approved the use of the database for research. The analysis using the eICU-CRD is exempt from institutional review board approval due to the retrospective design, lack of direct patient intervention, and the security schema, for which the re-identification risk was certified as meeting safe harbor standards by an independent privacy expert (Privacert, Cambridge, MA) (Health Insurance Portability and Accountability Act Certification no. 1031219-2). All experiments were performed in accordance with relevant guidelines and regulations.

**Consent for publication**

Not applicable.

**Availability of data and materials**

The datasets analyzed in the current study are publicly available in the MIMIC-III repository (https://mimic.physionet.org/) and eICU-CRD repository (https://eicu-crd.mit.edu/).

**Competing interests**

The authors declare that they have no competing interests.

**Funding**

LAC is funded by the National Institute of Health through NIBIB R01 EB017205.

**Authors' contributions**

LAC, VO and BM conceived the presented idea. VO and BM developed the theory and performed the computations. BM, LAMS, WY and DJS led the drafting of the manuscript. LAC and VO supervised the findings of this work. All authors discussed the results and contributed to the final manuscript.

**Acknowledgements**

Not applicable

# Additional files

Appendix 1 (.docx)
Sensitivity analysis with an increase of 10% in lactate levels

Appendix 2 (.docx)
Sensitivity analysis with fixed time difference between subsequent lactate measurements

Appendix 3 (.docx)
Patient characteristics, variable selection and cohort selection diagram

Appendix 4 (.docx)
Model development, hyperparameter tuning and optimization

Appendix 5 (.docx)
Overall performance metrics for the internal validation of MIMIC-III and eICU-CRD cohorts as well as external validation of the eICU-CRD cohort on the MIMIC-III cohort

Appendix 6 (.docx)
Calibration of each lactate prediction model per subgroup

Appendix 7 (.docx)
Strengthening the Reporting of Observational studies in Epidemiology (STROBE) Statement

# Appendix 1 – Sensitivity analysis with an increase of 10% in lactate levels

## Rationale

We performed this analysis to investigate whether a change of at least 10% in lactate levels influences the predictive performance of the model. As with the previous analysis, we defined the outcomes for each of the groups based on an increase or decrease of more than 10% in lactate levels as outlined in Table 4.

*Table 4 Alternative definition of lactate outcomes based on a 10% change in serum lactate levels.*

|  | Outcome | |
|---|---|---|
| **Initial Lactate Value** | Negative | Positive |
| Normal (< 2 mmol/L) | Serum lactate **increase by ≥ 10%** to ≥ 2 mmol/L | Serum lactate remains < 2 mmol/L |
| Mild (2 mmol/L to 4 mmol/L) | Serum lactate **increase by ≥ 10%** or any lactate increase resulting in a value > 4 mmol/L | Serum lactate stable, **decrease, or increase < 10%** (while remaining < 4 mmol/L) |
| Severe (> 4 mmol/L) | Serum lactate stable, **decrease by < 10%** or any increase in serum lactate | Serum lactate **decrease by ≥ 10%** |

## Cohort Design

We retrospectively evaluated a subgroup of adult patients (age ≥ 18 years) from the MIMIC-III dataset who had at least two serum lactate measurements recorded within the first 48 hours of ICU admission as well as an ICU length of stay ≥ 24 hours. The selected patients were further divided into three subgroups based on their initial serum lactate levels- i) normal group (< 2 mmol/L), ii), mild elevation group (2 mmol/L to 4 mmol/L), and iii) severe elevation group (> 4 mmol/L).

## Variable selection

We selected the same 54 variables available in the MIMIC-III database within 48 hours of admission, including admission diagnoses, medication, laboratory values, vital signs, patient demographics, and nursing care data obtained during the admission assessment as shown in Appendix 3. This also includes values obtained through an arterial blood gas (ABG) measurement which had to be sampled at least two hours (discretization interval) before the timestamp of the predicted serum lactate level.

# Results

The MIMIC-III study cohort had 29,337 serum lactate values recorded within the first 48 hours of ICU admission with close to half belonging to the normal group (46.9%) while the remaining values belonged to the mild (37.7%) and severe (15.4%) groups.

Positive outcomes with respect to serum lactate trajectory were observed in 87.4% (n= 12,026) of patients in the normal subgroup, 82% (n= 9,071) of patients in the mild subgroup, and 58.9% (n= 2,657) of patients in the severe subgroup. Figure 1 summarises the performance of each model for each subgroup. We also calculated precision-recall for different thresholds to devise curves that compare the performance of the models in presence of imbalanced datasets, as it is the case with the normal subgroup where positive outcomes (87.4%) significantly exceed negative outcomes (12.6%).

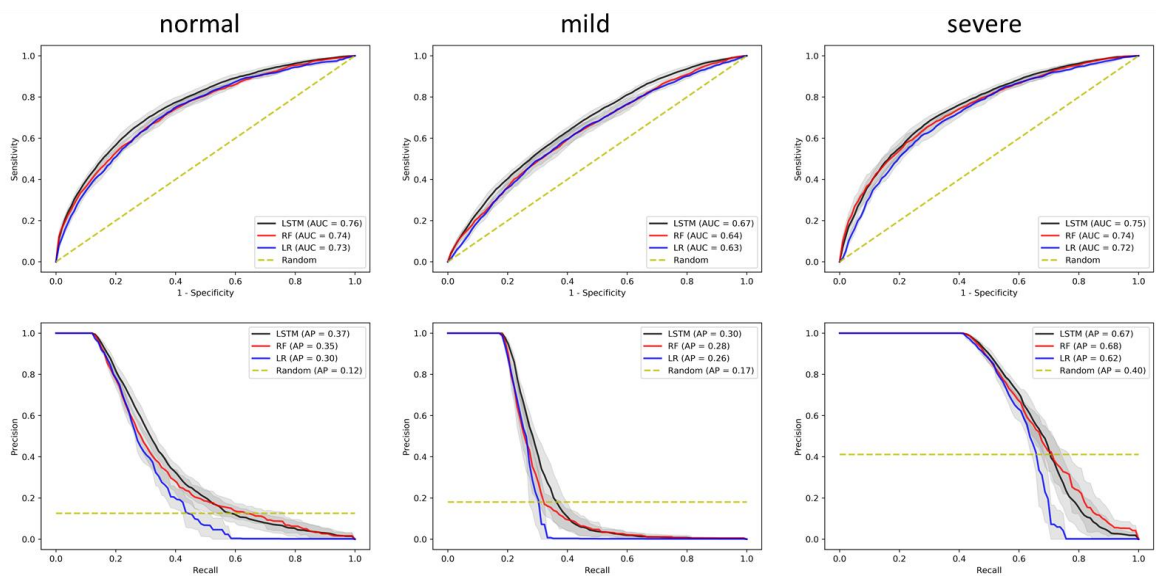

*Figure 5 Performance of each model in the MIMIC-III cohort across the three patient subgroups based on alternative definition of lactate outcomes. The top row represents AUC-ROC, while the bottom row shows the AU-PRC. Confidence intervals are shown in grey.*

Random Forest (RF) and Logistic Regression (LR) performed similarly for the normal and mild subgroup, while there was a higher difference in performance in the severe subgroup. The long short-term memory (LSTM) model on the other hand performed best across all three subgroups, achieving an AUC of 0.76 (95% CI 0.755-0.766) for the normal subgroup, 0.67 (95% CI 0.661-0.672) for the mild subgroup and 0.75 (95% CI 0.746-0.758) for the severe group.

These results are lower across all three subgroups in comparison to the model's performance with the definition of outcome based on changes between subgroups, where the LSTM model achieved an AUC of 0.77 (95% CI 0.762-0.771) for the normal subgroup, an AUC of 0.77 (95% CI 0.768-0.772) for the mild subgroup and an AUC of 0.85 (95% CI 0.840-0.851) for the severe group. It is notable that the performance differences in the mild and severe subgroups, with 18% and 10% difference in AUC, respectively, are also reflected in the AU-PRC.

# Appendix 2 – Sensitivity analysis with fixed time difference between subsequent lactate measurements

## Rationale

We performed this analysis to investigate whether a difference in time between two lactate measurements influences the prediction performance of lactate deterioration.

## Cohort Design

We retrospectively evaluated a subgroup of adult patients (age ≥ 18 years) from the MIMIC-III dataset that had at least 2 serum lactate measurements recorded within the first 48 hours of ICU admission as well as an ICU length of stay greater than or equal to 24 hours. Furthermore, we restricted the cohort to only those patients that had lactate measured subsequently within 8 hours of the preceding lactate measurement. The 8-hour interval was chosen based on the Surviving Sepsis Campaign guidelines, which recommend lactate to be measured every 6 hours. We allowed for a 2-hour delay as lactate levels might be measured but might not be immediately recorded in the patient's health record. The selected patient cohort was further divided into three subgroups based on their initial serum lactate levels- i) normal group (< 2 mmol/L), ii) mild elevation group (2 mmol/L to 4 mmol/L), and iii) severe elevation group (> 4 mmol/L). For each subgroup, we used between groups definition of outcomes as reported in the main text of the paper.

## Variable selection

We selected the same 54 variables available in the MIMIC-III database within 48 hours of admission, including admission diagnoses, medication, laboratory values, vital signs, patient demographics, and nursing care data obtained during the admission assessment as shown in Appendix 3. This also includes values obtained through an arterial blood gas (ABG) measurement which had to be sampled at least two hours (discretization interval) before the timestamp of the predicted serum lactate level.

## Results

The MIMIC-III study cohort had a decrease of serum lactate values recorded from 29,337 to 18,956 within the first 48 hours of ICU admission, with 41.1% belonging in the normal group, while the remaining values belonged to the mild (39.7%) and severe (19.2%) groups.
Positive outcomes with respect to serum lactate trajectory were observed in 85.6% (n=6,666) of patients in the normal subgroup, 33.5% (n=2,523) of patients in the mild subgroup, and 34.5% (n=1,253) of patients in the severe subgroup. Figure 6 summarizes the performance of each model for each subgroup. We also calculated precision-recall for different thresholds to devise the curves that compare the performance of the models in presence of imbalanced datasets, as is the case with the normal subgroup where positive outcomes (85.6%) significantly exceed negative outcomes (14.4%).

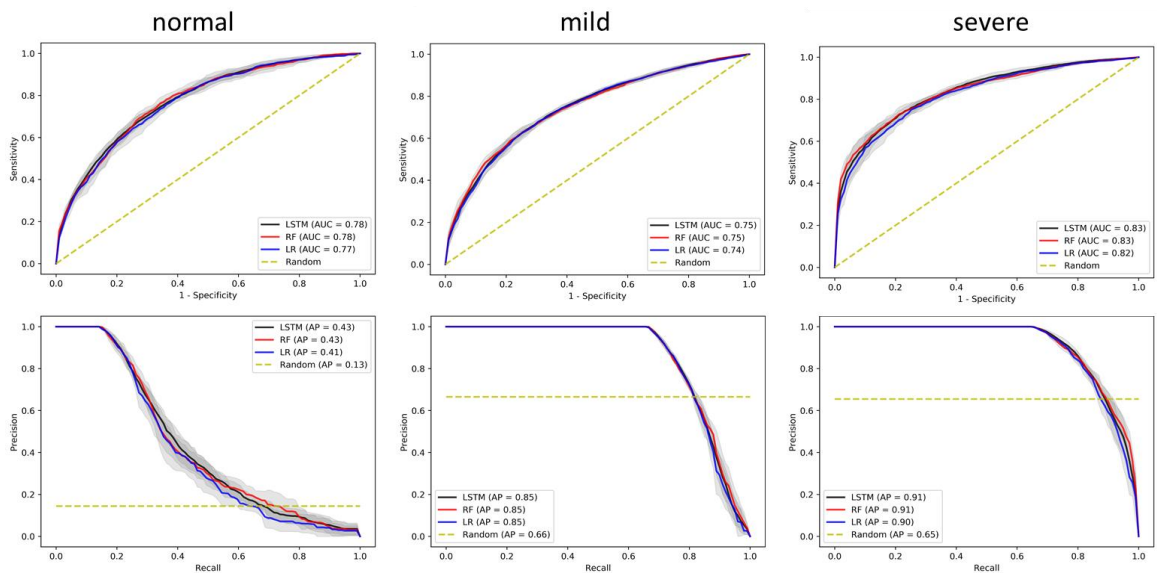

*Figure 6 Performance of each model in the MIMIC-III cohort across the three patient subgroups in the cohort restricted to only those patients that had the subsequent lactate measured within 8 hours of the preceding lactate measurement. Top row represents AUC-ROC, while the bottom AU-PRC. Confidence intervals are shown in grey.*

Random Forest (RF), Logistic Regression (LR) and long short-term memory (LSTM) performed similarly across the three subgroups, with LR showing a slightly lower performance (1%). Performance of the LSTM model on this cohort showed an AUC of 0.78 (95% CI 0.772-0.783) for the normal subgroup, an AUC of 0.75 (95% CI 0.744-0.751) for the mild subgroup, and an AUC of 0.83 (95% CI 0.831-0.840) for the severe group.

These results are lower across the mild and severe groups in comparison to the performance of the model without a restriction on the time difference between lactate measurements where the LSTM model achieved an AUC of 0.77 (95% CI 0.768-0.772) for the mild subgroup and an AUC of 0.85 (95% CI 0.840-0.851) for the severe group. In the normal subgroup, the performance increased by 1% from an AUC of 0.77 (95% CI 0.762-0.771).

# Appendix 3 – Patient characteristics, variable selection and cohort selection diagram

*Table 5 Selected variables grouped by type, common between MIMIC-III and eICU-CRD patient cohorts.*

| Type | Variables |
|---|---|
| Laboratory | Glucose, ALT, AST, Differential-Basos, Differential-Eos, Differential-Lymphs, Differential-Monos, BUN, Anion Gap, Base Excess, pH, Sodium, Potassium, Calcium, Magnesium, Phosphate, Bicarbonate, Creatinine, Chloride, Bilirubin, Red Blood Cells, White Blood Cells, Platelets, Hematocrit, Hemoglobin, Troponin T, RDW, Alkaline Phosphate, Ionized Calcium, PaCO2, PaO2, Lactate, MCV, MCH, MCHC, PT, PTT |
| Clinical Observations | Heart Rate, Respiratory Rate, ART BP mean, ART BP Diastolic, ART BP Systolic, Temperature, GCS-Total, GCS-Eye Opening, GCS-Motor Response, GCS-Verbal Response, Oxygen Saturation, CVP, Urine Output |
| Treatment | FiO2 |
| Demographic | Age, Weight, Gender |

*Table 6 Imputation strategy for each variable.*

| **Variables** | **Imputation strategy** |
|---|---|
| Glucose, ALT, AST, Anion Gap, Differential-Basos, Differential-Eos, Differential-Lymphs, Differential-Monos, BUN, Base Excess, RDW, pH, Sodium, Potassium, Calcium, Magnesium, Phosphate, Bicarbonate, Creatinine, Chloride, Bilirubin, Red Blood Cells, White Blood Cells, Platelets, Hematocrit, Hemoglobin, Troponin T, Alkaline Phosphate, Ionized Calcium, PaCO2, PaO2, Lactate, Heart Rate, Respiratory Rate, Temperature, Oxygen Saturation, CVP, Urine Output, FiO2, MCV, MCH, MCHC, PT, PTT | Forward Fill, then Median if missing completely |
| GCS - Eye Opening, GCS - Motor Response, GCS - Verbal Response, GCS-Total | Forward Fill, then Mode if missing completely (After calculating GCS Total using the other GCS aspects if available) |
| ART BP mean, ART BP Diastolic, ART BP Systolic | Combine with Non-Invasive Blood Pressure (NBP), Forward Fill, then Median if missing completely |

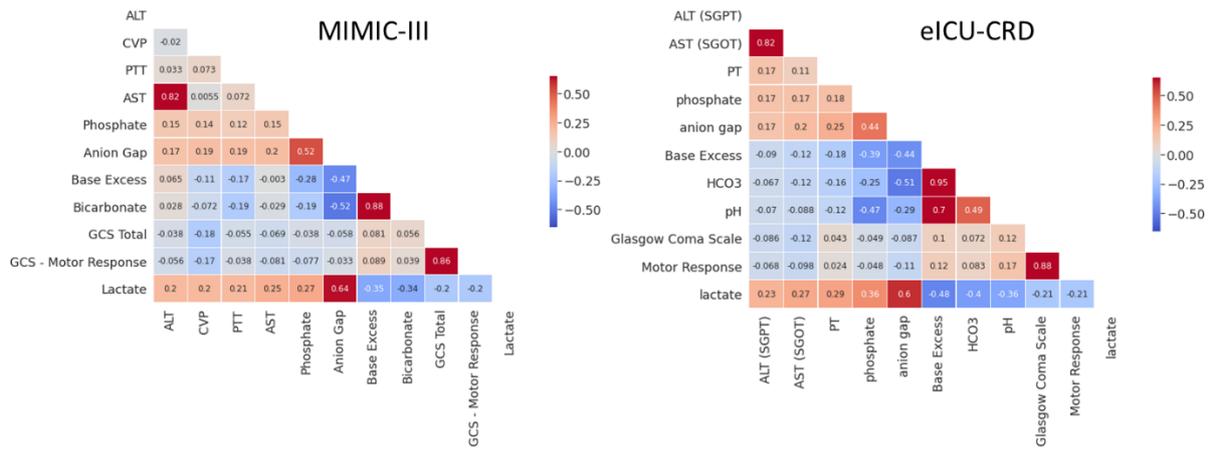

*Figure 7 Linear correlation (Pearson) between lactate and 10 highest correlated variables and lactate in both MIMIC-III and eICU-CRD patient cohorts.*

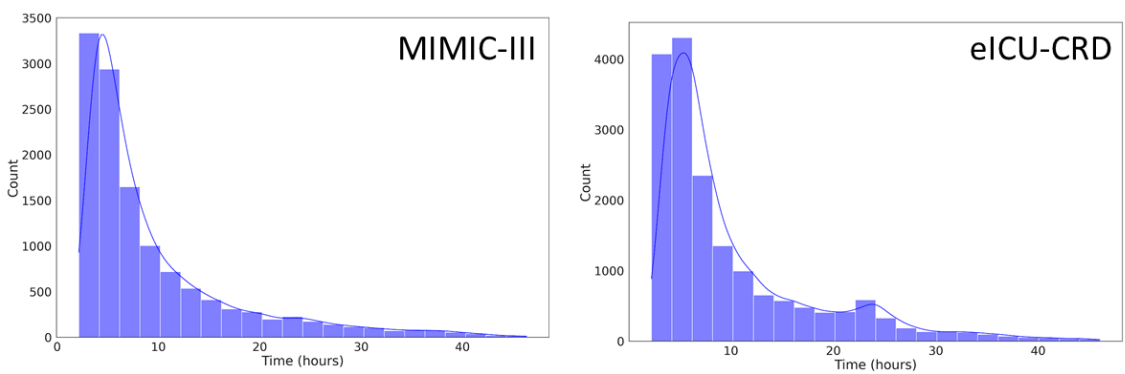

*Figure 8 Distribution of time difference between subsequent lactate measurements.*

*Table 7 MIMIC-III patient cohort characteristics across subgroups.*

| Variable | Normal subgroup (mean ± s.d.) | Mild subgroup (mean ± s.d.) | Severe subgroup (mean ± s.d.) |
|---|---|---|---|
| Demographic and ICU stay | | | |
| Age | 64.6 ± 16.5 | 64.4 ± 16.7 | 62.8 ± 17.0 |
| Sex (male) | 4502 (57%) | 3777 (58%) | 1380 (58%) |
| Length of stay (days) | 7.7 (± 8.6) | 7.9 (± 8.9) | 8.8 (± 9.5) |
| Mortality (%) | 16.6 | 21.5 | 37.9 |
| Laboratory | | | |
| Bicarbonate (mEq/L) | 22.7 ± 5.1 | 21.6 ± 4.5 | 19.0 ± 5.3 |
| Creatinine (mg/dL) | 1.8 ± 1.7 | 1.7 ± 1.5 | 2.0 ± 1.6 |
| Bilirubin (mg/dL) | 2.1 ± 4.0 | 3.5 ± 6.4 | 4.8 ± 7.0 |
| PaCO2 (mmHg) | 41.8 ± 11.3 | 39.1 ± 8.5 | 36.8 ± 9.1 |
| PaO2 (mmHg) | 133.7 ± 77.2 | 142.8 ± 82.8 | 138.0 ± 82.0 |
| BUN (mg/dL) | 32.8 ± 25.7 | 31.5 ± 23.4 | 33.5 ± 23.2 |
| pH | 7.4 ± 0.4 | 7.3 ± 0.4 | 7.3 ± 0.4 |
| Glucose (mg/dL) | 137.1 ± 52.2 | 145.7 ± 59.0 | 162.5 ± 84.6 |
| Anion Gap (mEq/L) | 13.6 ± 3.8 | 14.8 ± 4.0 | 20.0 ± 6.3 |
| Hematocrit (%) | 30.2 ± 5.0 | 30.7 ± 5.3 | 30.7 ± 5.7 |
| Hemoglobin (g/dL) | 10.3 ± 1.7 | 10.5 ± 1.9 | 10.5 ± 2.0 |
| Base Excess (mEq/L) | - 1.2 ± 4.8 | - 2.2 ± 4.2 | - 5.2 ± 5.8 |
| Phosphate (mg/dL) | 3.8 ± 1.6 | 3.8 ± 1.6 | 4.6 ± 2.1 |
| Clinical Observation | | | |
| Heart Rate (bpm) | 87.8 ± 18.2 | 91.3 ± 18.7 | 96.3 ± 19.6 |
| Respiratory Rate (insp/min) | 19.6 ± 5.8 | 19.9 ± 5.9 | 21.5 ± 6.5 |
| Oxygen Sat (%) | 97.1 ± 4.1 | 97.0 ± 4.8 | 96.1 ± 7.1 |
| Temperature (°C) | 37.1 ± 0.9 | 37.1 ± 0.9 | 36.9 ± 1.1 |
| GCS Total (points) | 10.0 ± 3.7 | 9.0 ± 4.0 | 7.6 ± 4.1 |
| Urine Output (ml/hr) | 162.7 ± 174.9 | 164.1 ± 191.7 | 149.3 ± 211.8 |
| Treatment | | | |
| FiO2 (%) | 53.5 ± 18.3 | 56.3 ± 19.4 | 61.0 ± 20.6 |

*Table 8 eICU-CRD patient cohort characteristics across subgroups.*

| Variable | Normal subgroup (mean ± s.d.) | Mild subgroup (mean ± s.d.) | Severe subgroup (mean ± s.d.) |
|---|---|---|---|
| Demographic and ICU stay | | | |
| Age | 61.9 ± 16.0 | 62.3 ± 16.3 | 61.8 ± 16.0 |
| Gender (M) | 5142 (54%) | 4772 (56%) | 2649 (55%) |
| Length of stay (days) | 5.6 (± 6.3) | 5.7 (± 6.6) | 6.1 (± 7.0) |
| Mortality (%) | 10.0 | 15.3 | 29.4 |
| Laboratory | | | |
| Bicarbonate (mEq/L) | 22.5 ± 5.7 | 21.3 ± 4.9 | 18.1 ± 5.4 |
| Creatinine (mg/dL) | 1.9 ± 1.9 | 1.9 ± 1.6 | 2.3 ± 1.6 |
| Bilirubin (mg/dL) | 1.3 ± 2.6 | 1.8 ± 3.6 | 2.6 ± 4.1 |
| PaCO2 (mmHg) | 42.0 ± 13.0 | 39.2 ± 11.0 | 36.8 ± 11.2 |
| PaO2 (mmHg) | 117.5 ± 70.3 | 122.2 ± 74.5 | 126.5 ± 80.5 |
| BUN (mg/dL) | 33.4 ± 26.3 | 34.2 ± 24.2 | 33.9 ± 23.2 |
| pH | 7.4 ± 0.1 | 7.3 ± 0.1 | 7.3 ± 0.1 |
| Glucose (mg/dL) | 144.8 ± 59.2 | 157.3 ± 69.9 | 171.3 ± 92.7 |
| Anion Gap (mEq/L) | 11.6 ± 4.9 | 12.6 ± 4.7 | 17.7 ± 6.7 |
| Hematocrit (%) | 30.6 ± 6.4 | 31.7 ± 7.1 | 31.1 ± 7.5 |
| Hemoglobin (g/dL) | 10.1 ± 2.2 | 10.5 ± 2.4 | 10.3 ± 2.5 |
| Base Excess (mEq/L) | - 1.8 ± 6.6 | - 3.4 ± 5.9 | - 8.1 ± 7.0 |
| Phosphate (mg/dL) | 3.5 ± 1.6 | 3.6 ± 1.6 | 4.6 ± 2.3 |
| Clinical Observation | | | |
| Heart Rate (bpm) | 89.6 ± 19.7 | 93.7 ± 20.1 | 97.7 ± 20.7 |
| Respiratory Rate (insp/min) | 19.7 ± 6.3 | 20.9 ± 6.5 | 22.3 ± 6.9 |
| Oxygen Sat (%) | 96.9 ± 3.3 | 96.8 ± 3.5 | 96.3 ± 5.3 |
| Temperature (°C) | 36.9 ± 1.2 | 36.9 ± 1.3 | 36.5 ± 1.6 |
| GCS Total (points) | 11.2 ± 3.9 | 10.7 ± 4.1 | 9.1 ± 4.4 |
| Urine Output (ml/hr) | 195.6 ± 386.6 | 175.3 ± 378.0 | 137.7 ± 280.4 |
| Treatment | | | |
| FiO2 (%) | 48.1 ± 21.3 | 51.2 ± 22.6 | 58.4 ± 25.5 |

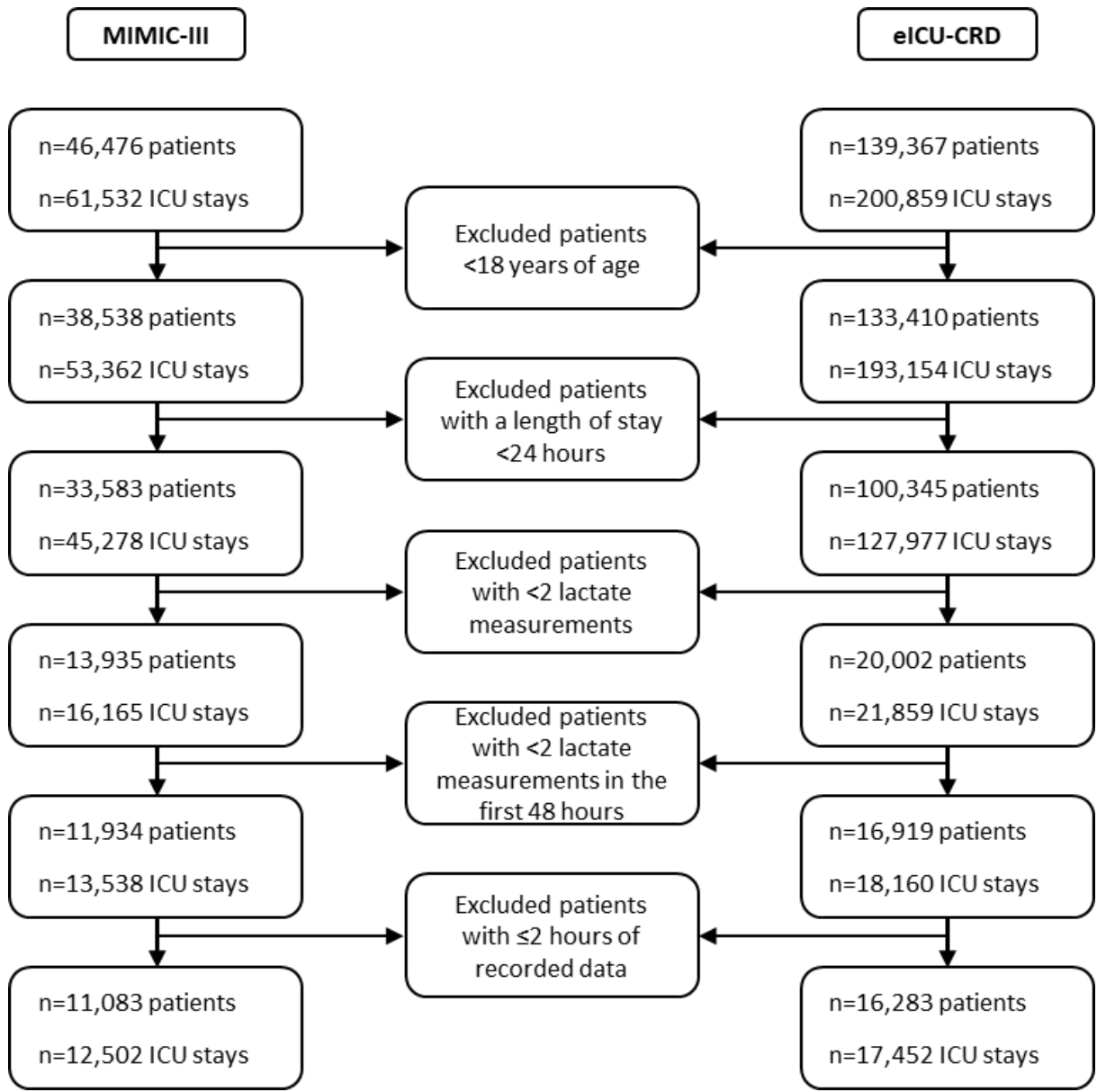

*Figure 9 Cohort selection diagram for both MIMIC-III and eICU-CRD datasets.*

# Appendix 4 – Model development, hyperparameter tuning and optimization

In our primary deep neural network models for all three subgroups, one layer of long short-term memory (LSTM) was initialized with Xavier normal weight initializer used to capture the time sequence in the data with 1024, 512, and 750 hidden neurons for the normal, mild, and severe groups, respectively. A Softmax function was applied on the last layer of the models to classify lactate data samples into the binary outcome of positive or negative. L1 regularization penalty and dropout of rate 0.5 were used to avoid overfitting of the models on the training datasets. All the LSTM models were trained with the Adam optimizer and learning rate of 0.0008 for 60 epochs with batch size of 64.

We dealt with imbalanced datasets in three steps. First, we applied stratified five-fold cross-validation to ensure that the class ratios are preserved after each train-test split. During the training process, the ratio of both classes was also maintained in each batch of data. Finally, class weights were added to the cross-entropy loss function to penalize the misclassification of the minority class.

To compare the performance of LSTM models in each subgroup for internal and external validation, Random Forest (RF) as a non-parametric and Logistic Regression (LR) as a parametric predictive machine learning model were developed. RF and LR were also optimized by randomized grid searching over their hyperparameters toward maximizing F-1 measure. Table 1 summarizes the final selected hyperparameters for all models used for each subgroup. All the models are implemented in Python using PyTorch and the Scikit-learn libraries.

*Table 1 Model hyperparameters across subgroups*

| Hyperparameter | **Normal group** | **Mild group** | **Severe group** |
|---|---|---|---|
| **LSTM** | | | |
| Hidden layer | 1 | 1 | 1 |
| Hidden neurons | 1024 | 512 | 750 |
| Dropout | 0.5 | 0.5 | 0.5 |
| L1 factor | 0.0004 | 0.0006 | 0.0005 |
| Optimization function | Adam (lr=0.0008) | Adam (lr=0.0008) | Adam (lr=0.0008) |
| Class weight | Inverse class ratio | Inverse class ratio | Inverse class ratio |
| **RF** | | | |
| Number of estimators | 500 | 400 | 300 |
| Criterion | Entropy | Entropy | Entropy |
| Max depth | 8 | 6 | 5 |
| Min samples split | 8 | 6 | 5 |
| Class weight | Inverse class ratio | Inverse class ratio | Inverse class ratio |
| **LR** | | | |
| Solver | liblinear | liblinear | liblinear |
| C | 1 | 1 | 1 |
| Class weight | Inverse class ratio | Inverse class ratio | Inverse class ratio |

# Appendix 5 – Overall performance metrics for the internal validation of MIMIC-III and eICU-CRD cohorts as well as external validation of the eICU-CRD cohort on the MIMIC-III cohort

For each model we detail the following performance metrics: Area Under the Receiver Operating Characteristic Curve (AUC), Average Precision (AP), Positive Predictive Value (PPV), Negative Predictive Value (NPV), harmonic mean of the precision and recall score (F-1) and Mathews Correlation Coefficient (MCC). 95% confidence intervals are shown in brackets for each algorithm, namely Logistic Regression (LR), Random Forest (RF) and long short-term memory (LSTM) neural network.

## MIMIC-III cohort
### Normal group

|      | AUC | AP | PPV | NPV | F-1 | MCC |
|------|-----|-----|-----|-----|-----|-----|
| **LR** | 0.74 (0.732-0.740) | 0.31 (0.298-0.312) | 0.34 (0.328-0.342) | 0.90 (0.903-0.904) | 0.35 (0.345-0.351) | 0.25 (0.242-0.251) |
| **RF** | 0.74 (0.738-0.748) | 0.36 (0.350-0.373) | 0.36 (0.353-0.369) | **0.91** (0.906-0.908) | 0.37 (0.367-0.376) | 0.28 (0.268-0.281) |
| **LSTM** | **0.77** (0.762-0.771) | **0.37** (0.358-0.377) | **0.38** (0.372-0.388) | **0.91** (0.907-0.909) | **0.38** (0.375-0.388) | **0.29** (0.282-0.297) |

### Mild group

|      | AUC | AP | PPV | NPV | F-1 | MCC |
|------|-----|-----|-----|-----|-----|-----|
| **LR** | 0.73 (0.722-0.730) | 0.77 (0.761-0.770) | 0.73 (0.730-0.735) | 0.61 (0.607-0.617) | 0.73 (0.732-0.739) | 0.34 (0.338-0.350) |
| **RF** | 0.75 (0.745-0.752) | 0.82 (0.815-0.820) | 0.73 (0.725-0.735) | 0.61 (0.605-0.617) | 0.74 (0.732-0.740) | 0.34 (0.329-0.349) |
| **LSTM** | **0.77** (0.768-0.772) | **0.83** (0.822-0.830) | **0.76** (0.753-0.761) | **0.63** (0.630-0.639) | **0.75** (0.747-0.753) | **0.40** (0.388-0.401) |

### Severe group

|      | AUC | AP | PPV | NPV | F-1 | MCC |
|------|-----|-----|-----|-----|-----|-----|
| **LR** | 0.82 (0.812-0.820) | 0.86 (0.858-0.867) | 0.80 (0.792-0.804) | 0.67 (0.665-0.679) | 0.79 (0.782-0.792) | 0.47 (0.462-0.484) |
| **RF** | 0.83 (0.832-0.838) | 0.89 (0.890-0.896) | **0.81** (0.812-0.818) | 0.67 (0.671-0.680) | 0.79 (0.788-0.795) | 0.49 (0.490-0.503) |
| **LSTM** | **0.85** (0.840-0.851) | **0.90** (0.891-0.901) | **0.81** (0.802-0.815) | **0.71** (0.679-0.721) | **0.81** (0.802-0.814) | **0.52** (0.504-0.528) |

## eICU-CRD cohort
### Normal group

|      | AUC | AP | PPV | NPV | F-1 | MCC |
|------|-----|-----|-----|-----|-----|-----|
| **LR** | 0.68 (0.671-0.683) | 0.23 (0.221-0.236) | 0.26 (0.254-0.275) | 0.89 (0.889-0.892) | 0.25 (0.242-0.268) | 0.15 (0.138-0.163) |
| RF | 0.70 (0.686-0.703) | 0.26 (0.250-0.260) | 0.28 (0.277-0.289) | 0.89 (0.891-0.894) | 0.27 (0.257-0.277) | 0.17 (0.161-0.175) |
| **LSTM** | **0.72** (0.707-0.724) | **0.27** (0.263-0.282) | **0.30** (0.293-0.313) | **0.90** (0.896-0.901) | **0.31** (0.295-0.316) | **0.20** (0.193-0.213) |

Mild group

|  | AUC | AP | PPV | NPV | F-1 | MCC |
|---|---|---|---|---|---|---|
| **LR** | 0.69 (0.683-0.693) | 0.74 (0.730-0.742) | 0.71 (0.703-0.713) | 0.59 (0.566-0.580) | 0.71 (0.705-0.715) | 0.28 (0.269-0.292) |
| **RF** | 0.73 (0.723-0.731) | **0.80** (0.798-0.802) | 0.72 (0.711-0.726) | 0.60 (0.587-0.607) | 0.73 (0.719-0.732) | 0.31 (0.302-0.323) |
| **LSTM** | **0.74** (0.735-0.745) | **0.80** (0.799-0.804) | **0.73** (0.728-0.738) | **0.61** (0.603-0.625) | **0.74** (0.730-0.744) | **0.35** (0.335-0.357) |

Severe group

|  | AUC | AP | PPV | NPV | F-1 | MCC |
|---|---|---|---|---|---|---|
| **LR** | 0.82 (0.812-0.821) | 0.88 (0.872-0.880) | 0.81 (0.808-0.814) | 0.66 (0.658-0.671) | 0.81 (0.808-0.814) | 0.47 (0.468-0.484) |
| **RF** | 0.83 (0.826-0.836) | 0.90 (0.897-0.906) | 0.81 (0.802-0.814) | 0.66 (0.656-0.672) | 0.81 (0.806-0.814) | 0.47 (0.460-0.480) |
| **LSTM** | **0.84** (0.837-0.848) | **0.91** (0.902-0.911) | **0.82** (0.814-0.824) | **0.68** (0.675-0.691) | **0.82** (0.817-0.825) | **0.50** (0.491-0.510) |

# External validation of eICU-CRD model on the MIMIC-III cohort

Normal group

|  | AUC | AP | PPV | NPV | F-1 | MCC |
|---|---|---|---|---|---|---|
| **LR** | 0.61 | 0.18 | 0.20 | 0.88 | 0.20 | 0.10 |
| **RF** | 0.69 | 0.24 | 0.25 | **0.89** | 0.26 | 0.15 |
| **LSTM** | **0.70** | **0.25** | **0.28** | **0.89** | **0.29** | **0.18** |

Mild group

|  | AUC | AP | PPV | NPV | F-1 | MCC |
|---|---|---|---|---|---|---|
| **LR** | 0.63 | 0.67 | 0.68 | 0.53 | 0.67 | 0.20 |
| **RF** | 0.74 | 0.81 | 0.72 | 0.60 | 0.73 | 0.32 |
| **LSTM** | **0.74** | **0.77** | **0.74** | **0.62** | **0.74** | **0.36** |

Severe group

|  | AUC | AP | PPV | NPV | F-1 | MCC |
|---|---|---|---|---|---|---|
| **LR** | 0.76 | 0.79 | 0.76 | 0.64 | 0.76 | 0.40 |
| **RF** | **0.83** | **0.89** | 0.79 | 0.69 | **0.80** | 0.48 |
| **LSTM** | **0.83** | 0.88 | **0.80** | **0.70** | **0.80** | **0.50** |

# Appendix 6 – Calibration of each lactate prediction model per subgroup

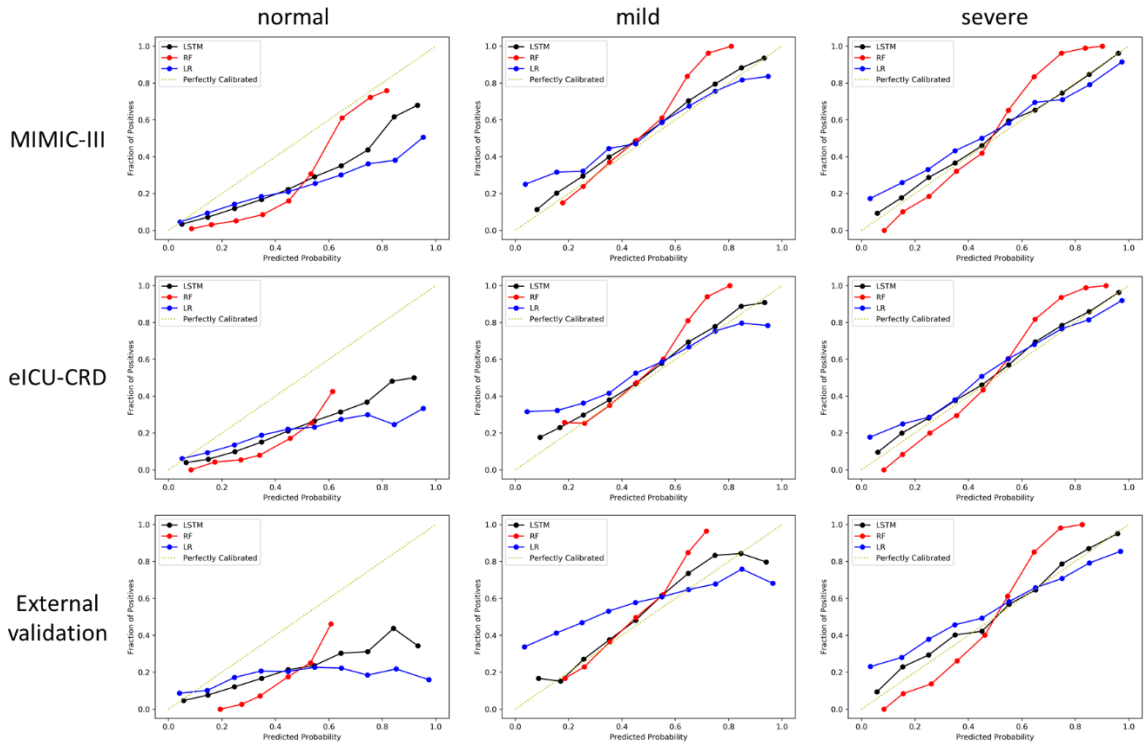

*Figure 10 Calibration curve for each patient subgroup and patient cohort, including external validation*